\documentclass[10pt,twocolumn,showpacs,amsmath,amssymb,osajnl,floatfix,superscriptaddress]{revtex4-1}

\usepackage{graphicx}
\usepackage{dcolumn}
\usepackage{bm}
\usepackage{color}
\usepackage{txfonts}
\usepackage{microtype}
\usepackage{overpic}

\begin{document}
\author{Yan-Fei Li}	\affiliation{School of Science, Xi'an Jiaotong University, Xi'an 710049, China}
\author{Jian-Xing Li}\email{jianxing@xjtu.edu.cn}
\affiliation{School of Science, Xi'an Jiaotong University, Xi'an 710049, China}
\author{Karen Z. Hatsagortsyan}
\email{k.hatsagortsyan@mpi-hd.mpg.de}
\affiliation{Max-Planck-Institut f\"{u}r Kernphysik, Saupfercheckweg 1,
	69117 Heidelberg, Germany}
\author{Yong-Tao Zhao}\email{zhaoyongtao@xjtu.edu.cn}
\affiliation{School of Science, Xi'an Jiaotong University, Xi'an 710049, China}	
\author{Bo Zhang}	\affiliation{Department of high energy density physics, Research center of laser fusion, 621900, Mianyang, Sichuan, China}
\affiliation{Science and technology on plasma physics laboratory, Research center of laser fusion, 621900, Mianyang, Sichuan, China}
\author{Yu-Tong Li}	\affiliation{Beijing National Laboratory for Condensed Matter Physics, Institute of Physics, Chinese Academy of Sciences, Beijing, 100190, China}\affiliation{School of Physical Sciences, University of Chinese Academy of Sciences, Beijing, 100049, China}
\author{Yang-Yang Liu}	\affiliation{School of Science, Xi'an Jiaotong University, Xi'an 710049, China}
\author{Ze-Long Zhang}	\affiliation{School of Science, Xi'an Jiaotong University, Xi'an 710049, China}
\author{Zhong-Feng Xu}	\affiliation{School of Science, Xi'an Jiaotong University, Xi'an 710049, China}
\author{Christoph H. Keitel}
\affiliation{Max-Planck-Institut f\"{u}r Kernphysik, Saupfercheckweg 1,
	69117 Heidelberg, Germany}	
	
	\bibliographystyle{apsrev4-1}

	\title{Determining Carrier-Envelope Phase of Relativistic Laser Pulses via \\Electron Momentum Distribution}
	\date{\today}

\begin{abstract}
	The impacts of the carrier-envelope phase (CEP) of a long relativistic tightly-focused laser pulse on the dynamics of a counter-propagating electron beam have been investigated in the, so-called, electron reflection regime,   
requiring the Lorentz factor of the electron $\gamma$  to be approximately two orders of magnitudes lower than the dimensionless laser field parameter $\xi$. 
The electrons are reflected at the rising edge of the laser pulse due to the ponderomotive force of the focused laser beam, and an asymmetric electron angular distribution  emerges along the laser polarization direction, which sensitively depends on the CEP of the driving laser pulse for weak radiative stochastic effects. 
The CEP siganatures are observable at laser intensities of the order or larger than $10^{19}$ W/cm$^2$ and the pulse duration  up to 10 cycles. The CEP detection resolution is proportional to the electron beam density and
can achieve approximately $0.1^{\circ}$  at an electron density of about $10^{15}$ cm$^{-3}$.
The method is applicable for currently available ultraintense laser facilities with the laser peak power from tens of terawatt to multi-petawatt region.	
\end{abstract}

\maketitle

In the last decade remarkable progress has been achieved in laser technique pushing the limits of the chirped pulse amplification method \cite{CPA}.
With short laser pulses a large peak power is reached at relatively low pulse energies, allowing development of
terawatt lasers and large scale petawatt facilities, and paving a way for relativistic laser pulses, nowadays available up to a peak intensity of $10^{22}$ W/cm$^2$
\cite{Vulcan,ELI,Exawatt,Yanovsky2008}.
Intense lasers have many applications, e.g., for laser electron or ion acceleration \cite{Mourou2006,Esarey2009,Macchi2013}, x- or $\gamma$-ray radiation sources \cite{Murnane1991,Corde2013,Schwoerer2006,Yu2013}, inertial confinement fusion  \cite{Tabak1994,Lindl2004,Park2014}, and for laboratory astrophysics \cite{Ryutov1999Similarity,Li2018Laboratory,Remington2006}.

The carrier-envelope phase (CEP) of a laser pulse is a crucial parameter to characterize the waveform of the field,
which significantly  affects the laser-matter interaction as in the nonrelativistic regime,
such as in strong field ionization \cite{Liu2004,Rathje2013,Sansone2010,Mucke2002}, and high-order harmonic generation (HHG) \cite{Christov1997,Ishii2014}, as well as in the relativistic regime, e.g., in
nonlinear Compton scattering \cite{Boca2009,Machenroth2010,Seipt2013}, and in the electron-positron pair production processes \cite{Hebenstreit2009,Titov2016,Krajewska2012,Nuriman2013,Jansen2016,Meuren2016}. However, as already was pointed out in \cite{Li2018}, the CEP effect can have a different character in the nonrelativistic and relativistic regimes. While in the nonrelativistic regime the CEP effect mostly depends on the asymmetry of the field around the laser pulse peak, in the relativistic regime some CEP effects are due to field asymmetry in the rising edge of the laser pulse and can be conspicuous even in multicycle pulses \cite{Li2018}.  Consequently, the known nonrelativstic methods for CEP measurement of few-cycle pulses, such as the stereographic above-threshold ionization (ATI) \cite{Paulus2001},
the $f-2f$ interferometry \cite{Holzwarth2000}, and the streaking methods \cite{Streaking},
are inapplicable for multicycle laser beams in the relativistic regime.
For relativistic laser pulses, the angular distribution of the electron radiation \cite{Machenroth2010}, and the differential cross sections of the Breit-Wheeler pair production process \cite{Titov2016} are theoritically proposed to detect the CEPs of ultra-short laser pulses (the pulse duration no more than 2 cycles). However, the laser pulse duration for realistic relativistic laser pulses is usually longer than 15 fs ($\sim$ 6 cycles) \cite{Tajima1979,Mangles2004,Faure2004,Esarey2009,Leemans2006GeV,Hegelich2006,Mourou2006,Macchi2013,Vulcan,ELI,Exawatt,Yanovsky2008}. In our previous work \cite{Li2018} we have shown that the CEP determination of long laser pulses is possible by employing the fine features of the electron-radiation x-ray spectra. The method, however, is applicable only for extreme laser intensities $I_0 \gtrsim 10^{22}$ W/cm$^{-2}$, keeping open the problem of the CEP measurement of laser pulses in an intensity range from $10^{19}$ to $10^{22}$ W/cm$^2$, mostly produced in current ultraintense laser facilities.

\begin{figure}[t]
	\includegraphics[width=0.95\linewidth]{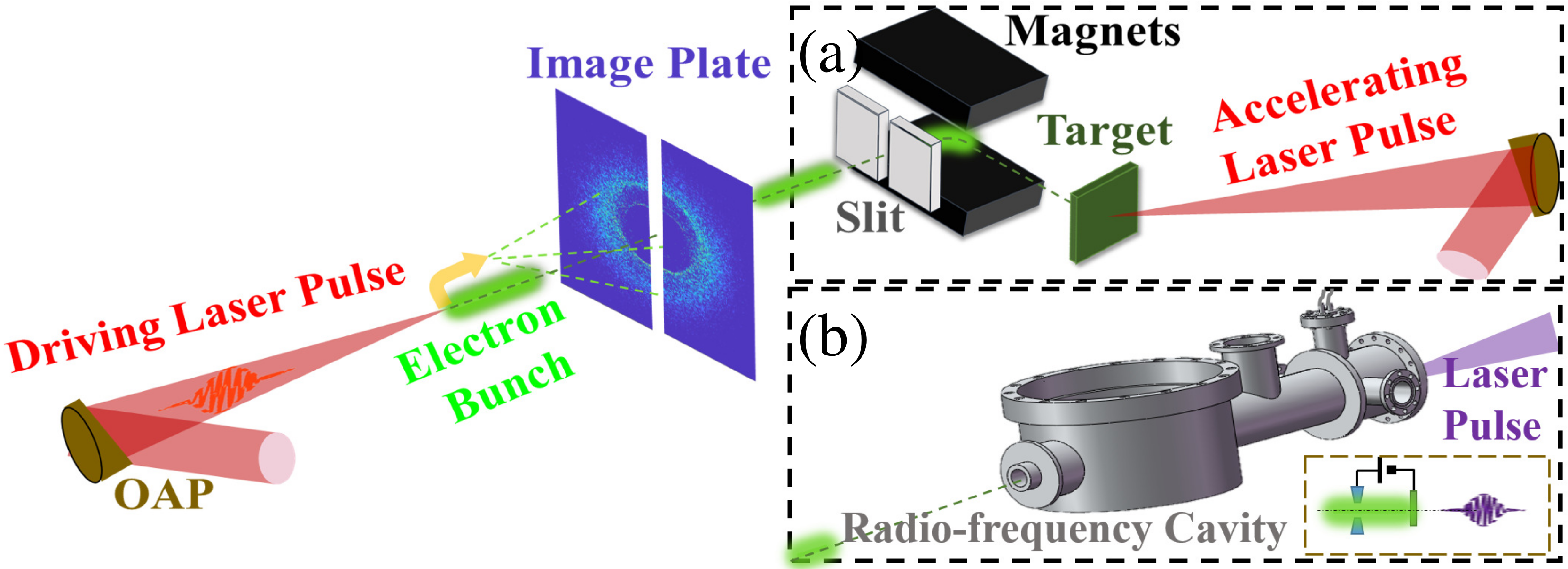}
	\caption{ The scenario of the CEP detection via the electron momentum distribution. The driving laser pulse is focused by an off-axis paraboloid (OAP) and collides head-on with an electron bunch in the $\gamma\ll\xi/2$ regime.
The bunch is produced either by (a) laser-plasma accelerator (an accelerating laser pulse ionizes the low-density target and accelerates the electrons forwards, and the magnets accompanied with a slit are used to select certain-energy electrons),  or (b)  radio-frequency (RF)
electron gun (a high-power laser pulse illuminates a photocathode surface placed on an end wall of a RF cavity, and the emitted electrons are accelerated immediately to a relativistic energy by the strong RF field in the cavity).  Due to the laser focusing effect, the electrons are reflected by the rising edge of the driving laser and move backwards. On the image plate, an asymmetric transverse momentum distribution of  electrons can be observed, which sensitively depends on the CEP of the driving laser pulse.
	}
	\label{fig1}
\end{figure}

  In this letter, we investigate the CEP determination of relativistic multicycle laser pulses with peak intensities  $I_0 \gtrsim 10^{19}$ W/cm$^{2}$, exploiting the momentum distribution of the electron beam after the interaction.
  The relativistic laser pulse interacts with a counterpropagating electron beam.
  The electron energy is considered to be much smaller than the reflection condition \cite{Piazza2012},
   i.e., $\gamma\ll\xi/2$, with the electron Lorentz factor $\gamma$,
      and the dimensionless parameter of the laser field $\xi\equiv e E_0/(m \omega_0)$. Here, $E_0$ and $\omega_0$ are the amplitude and frequency of the laser field, respectively, and $-e$ and $m$ the electron charge and mass, respectively. Planck units $\hbar=c=1$ are used throughout. Due to the laser focusing effect, the electrons are reflected by the rising edge of the driving laser pulse and move backwards, see the interaction scenario in Fig.~\ref{fig1}.  The electron beam can be  generated by either the laser-plasma accelerators \cite{Esarey2009,Wang2018}, or the radio-frequency (RF) electron gun system \cite{Maxson2017,Qi2015,Zhu2015}. We choose conditions when      the stochastic effects in the electron radiation are weak, i.e., the invariant quantum parameter $\chi\equiv |e|\sqrt{(F_{\mu\nu}p^{\nu})^2}/m^3\ll1$  \cite{Howard1962,Ritus1985}, where $F_{\mu\nu}$ is the field tensor, and $p^{\nu}=(\varepsilon,\textbf{p})$  the incoming electron 4-momentum. In this case the electron final transverse momentum distribution is asymmetric. The asymmetry  sensitively depends on  the CEP of the driving laser pulse and can be employed for CEP determination.
        This method does not rely on the electron radiation and, therefore, is applicable at
       much lower laser intensities ($\xi\gtrsim 5 $) than the CEP determination via the x-ray spectra ($\xi\gtrsim 100$) \cite{Li2018}. Meanwhile, the resolution of the CEP determination in this method is much higher than that in \cite{Li2018}, since the diffraction limitation of an electron is much smaller than that of a photon. The asymmetry due to CEP is larger in the rising edge of the laser pulse than near the peak, which allows the application of the method for rather long laser pulses ($\gtrsim 10$ cycles, $\sim 30-40 fs$) and even for longer flat-top laser pulses ($\gtrsim 20$ cycles). 

In relativistic regime, as considered  $ I_0\gtrsim 10^{19}$ W/cm$^2$ ($\xi\gtrsim5$), the quantum radiation-reaction effects could still not be negligible, since $\chi\sim 10^{-6} \xi\gamma$. To keep the consistency of the  simulations,
we carry out the calculation of the radiation based on Monte-Carlo approaches employing QED theory for the electron radiation and classical equations of motion for the propagation of electrons between photon emissions \cite{Elkina2011,Ridgers2014Modelling,Green2015}. The photon emission probability in the local constant field approximation is used, which is determined by the local value of the parameter $\chi$ and accounts for the quantum
recoil due to photon emission. As $\chi\ll1$, the quantum stochastic and recoil effects at photon emission are rather weak \cite{Shen1972,Duclous2011,Li2018,Neitz2013}, and the electron dynamics can be also described by Landau-Lifshitz equation
\cite{Landau1975,Zhidkov2002}
and gives similar results \cite{supplemental}. When $\chi\lesssim 10^{-3}$, the radiation-reaction effects are negligible \cite{Piazza2012}, and the electron dynamics is well described by Newton equation with the Lorentz force \cite{supplemental}.

We consider a  linearly polarized and tightly focused laser pulse with a Gaussian temporal profile propagating along $+z$-direction and  polarized in $x$-direction: $\bf{E}(\bf{r}, \eta)\varpropto$exp[$i(\eta+\psi_{\rm{CEP}})$]exp($-t^2/\tau^2$), with
$\eta=\omega_0t-k_0z$, the CEP $\psi_{\rm{CEP}} $ \cite{supplemental}, the laser wavevector $k_0=2\pi/\lambda_0$, the wavelength $\lambda_0$, and the pulse duration $\tau$.
The spatial distribution of the fields takes into account up to the $\epsilon^3$-order of the nonparaxial corrections \cite{Salamin2002,Salamin2002PhysRevSTAB}, where $\epsilon=w_0/z_r$, $w_0$ is the laser beam radius at the focus, and $z_r=k_0 w_0^2/2$ the Rayleigh length.

 \begin{figure}[t]
	\includegraphics[width=0.9\linewidth]{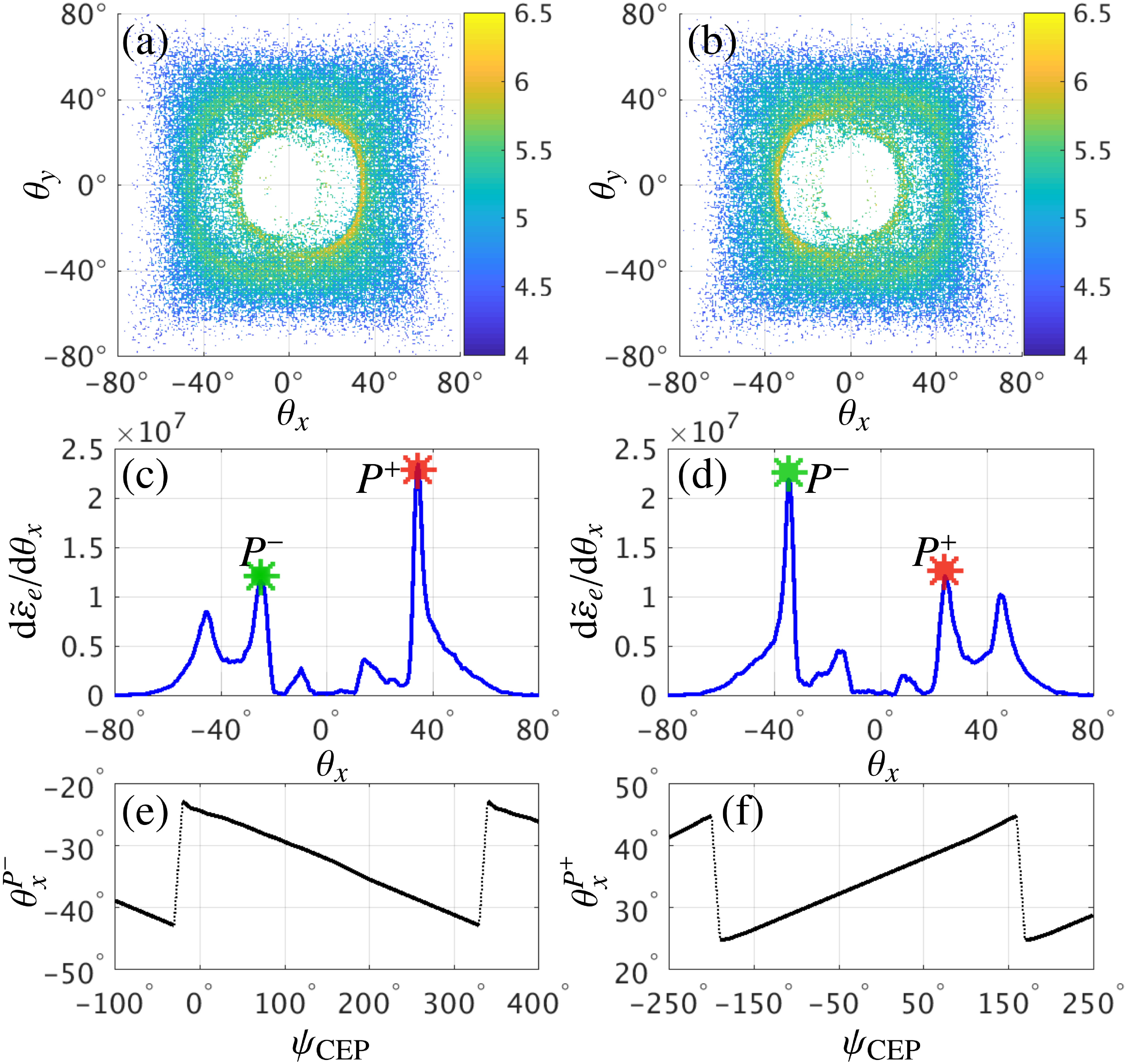}
	\caption{\label{fig2} (a), (b) Angle-resolved electron energy  log$_{10}$[d$^2\varepsilon_e$/(d$\theta_x$d$\theta_y$)] rad$^{-2}$ in units of $m$ vs the transverse deflection angles of the electron momenta
	 $\theta_x=$ arctan($p_x/p_z$) and $\theta_y=$ arctan($p_y/p_z$) with the CEP $\psi_{\rm CEP}=0^{\circ}$ and $180^{\circ}$, respectively. (c), (d) d$\tilde{\varepsilon}_e$/d$\theta_x$
	  vs $\theta_x$
	  with $\psi_{\rm CEP}$ = 0$^\circ$ and 180$^\circ$, respectively. Here, d$\tilde{\varepsilon}_e$/d$\theta_x$ = $\int_{-10^{\circ}}^{10^{\circ}}$d$^2\varepsilon_e$/[d$\theta_x$d$\theta_y$] d$\theta_y$: the total angle-resolved electron energy in the angle region of $-10^{\circ}\leq\theta_y\leq10^{\circ}$.	And, the peaks in the regions of $\theta_x<0^{\circ}$ and $\theta_x>0^{\circ}$ are marked by the green-star $P^{-}$ and the red-star $P^{+}$, respectively.
	    (e), (f) The variations of $\theta_x^{P^{-}}$ and $\theta_x^{P^{+}}$ corresponding to $\theta_x$ of $P^{-}$ and $P^{+}$ with respect to $\psi_{\rm CEP}$, respectively. The employed laser and electron parameters are given in the text.}	
\end{figure}

 A typical example of the angle-resolved electron momentum distribution in the considered regime, which will be employed further for the CEP determination,  is presented in Fig.~\ref{fig2}. The peak intensity of the laser pulse is $I_0\approx5\times10^{20}$ W/cm$^2$ ($\xi=20$), $\lambda_0=1$ $\mu$m, $\tau = 6T_0$, $T_0$ is the laser period and $w_0=2\,\mu$m.
An electron bunch of a cylindrical form collides head-on with the laser pulse at the polar angle $\theta_e=179^{\circ}$ and the azimuthal angle $\phi_e=0^{\circ}$. The electron mean initial kinetic energy is $\varepsilon_i=0.1$ MeV ($\gamma\approx0.2$, and $\chi_{max}\approx 3\times10^{-5}$), the electron bunch radius  $w_e= 2\lambda_0$, the length $L_e = 6\lambda_0$ and the density $n_e\approx 10^{15}$ cm$^{-3}$. The energy and angular spreads are $\Delta \varepsilon_i/\varepsilon_i =0.05$ and $\Delta \theta =0.02$, respectively (at larger energy spread $\Delta \varepsilon_i/\varepsilon_i =0.1$
or larger kinetic energies comparable results are obtained \cite{supplemental}).  The electrons in the bunch have a Gaussian distribution in the transverse direction and a uniform  distribution in the longitudinal one. This kind of electron bunch can be obtained by tightly focusing a laser pulse on the cathode of a RF photoinjector \cite{Maxson2017}.

 \begin{figure}[t]
 	\includegraphics[width=0.9\linewidth]{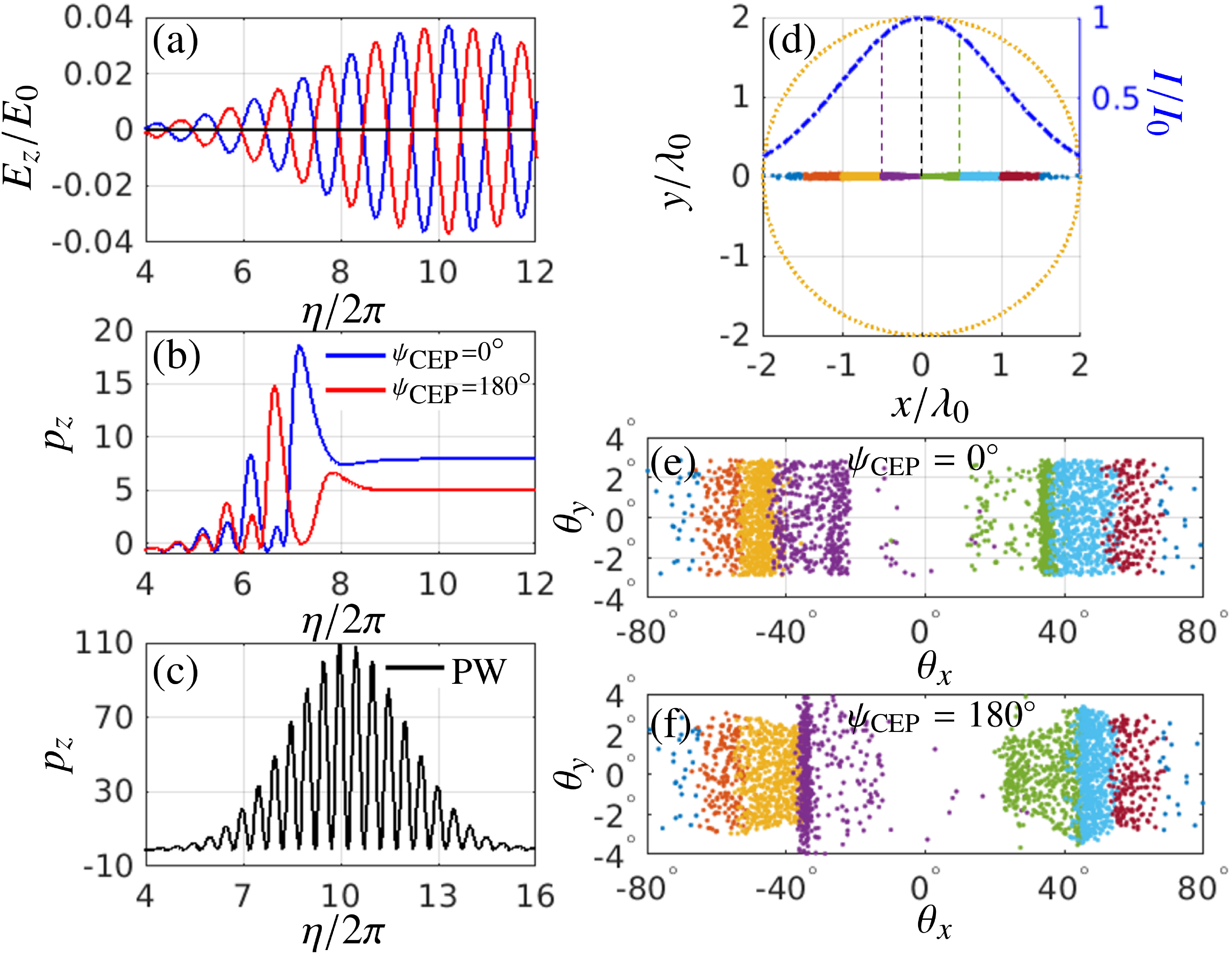}
	\caption{(a) The variations of the longitudinal component of electric fields $E_z$ at the position of (0.5$\lambda_0$, 0, 0) with respect to the laser phase $\eta$. The blue-solid and red-solid curves represent the focused laser fields with $\psi_{\rm CEP}=0^{\circ}$ and $180^{\circ}$, respectively. And, the black-solid curve represents  a plane wave with a 6-cycle Gaussian envelope.  The longitudinal momentum $p_z$ of a sample electron propagating in (b) the focused laser fields with $\psi_{\rm CEP}=0^{\circ}$ (blue) and $180^{\circ}$ (red), respectively, and (c) the plane wave (black) with respect to $\eta$.  (d) The initial transverse coordinate distribution of the sample electrons near the laser polarization plane ($y=0$), and the yellow circle shows the boundary of the electron bunch. The blue-dash-dotted curve represents the transverse Gaussian profile of the laser intensity $I$ scaled by the peak intensity $I_0$. (e), (f) The final transverse momentum distributions of the samle electrons of (d) in the focused laser fields with    $\psi_{\rm CEP}=0^{\circ}$ and $180^{\circ}$, respectively. In (d)-(f) the same color represents the same group of sample electrons. Other laser and electron parameters are the same as in Fig.~\ref{fig2}.  }
\label{fig3}
\end{figure}

Since $\gamma\ll\xi/2$, the electrons are reflected  and move backwards due to the the ponderomotive force caused by the laser focusing  (the condition of $w_0\approx w_e\sim \lambda_0$ is required to ensure the laser focusing effects are important, see details in \cite{supplemental}). 

 During the reflection process, the electron dynamics is sensitively governed  by the laser field, and, consequently, the field structural information, in particular CEP,   is encoded in the final electron momentum distribution, see  Fig.~\ref{fig2}. With different CEPs the deflection angle of the electron momenta $\theta_x=$ arctan($p_x/p_z$) with respect to the laser polarization direction, is very different. For the cases of $\psi_{\rm CEP}=0^{\circ}$ and $180^{\circ}$, the electron energy distribution is almost symmetric, and the slight deviation from the symmetry stems from the head-on colliding angle of the electron bunch $\theta_e=179^{\circ}$ (not exact $180^{\circ}$ for the experimental feasibility). For inspecting the main asymmetry features, the electron distributions in Figs.~\ref{fig2}(a) and \ref{fig2}(b) are integrated over $\theta_y$ from $-10^{\circ}$ to $10^{\circ}$, see Figs.~\ref{fig2}(c) and \ref{fig2}(d), respectively. In the regions of $\theta_x<0^{\circ}$ and $\theta_x>0^{\circ}$, the angle-resolved electron energy distribution d$\tilde{\varepsilon}_e$/d$\theta_x$ has two peaks indicated by $P^{-}$ (green star) and $P^{+}$ (red star). The angles $\theta_x$ corresponding to $P^{-}$ and $P^{+}$ are denoted by $\theta_x^{P^{-}}$ and  $\theta_x^{P^{+}}$, respectively. As the CEP varies within one period, $\theta_x^{P^{-}}$ ($\theta_x^{P^{+}}$) monotonously decreases (increases) approximately by 20.05$^{\circ}$ (20.05$^{\circ}$), namely from $-22.92^{\circ}$ to $-42.97^{\circ}$ (from $24.64^{\circ}$ to $44.69^{\circ}$), see Figs.~\ref{fig2}(e) and \ref{fig2}(f). Taking into account that an angular resolution less than 0.1 mrad is achievable with current technique of electron detectors \cite{Wang2013,Leemans2014,Wolter2016,Chatelain2014}, we may conclude that the CEP resolution here could reach $\sim0.1^\circ$. Since the resolution of the electron detector is proportional to the electron density, therefore, a higher CEP resolution is feasible with a higher density electron bunch.
        
The CEP signatures on electron dynamics are analyzed in Fig.~\ref{fig3}. Firstly, the $E_z$ components of the focused laser fields with $\psi_{\rm CEP}=0^{\circ}$ (blue) and $180^{\circ}$ (red), respectively, and the plane wave with a 6-cycle Gaussian envelope (black)  at the point of ($0.5\lambda_0$, 0, 0) are compared in Fig.~\ref{fig3}(a). The corresponding dynamics of the longitudinal momentum $p_z$ of a sample electron propagating in those three laser fields are illustrated in Figs.~\ref{fig3}(b) and \ref{fig3}(c). For intuitive understanding let us neglect for a moment radiation reaction, which in fact has a minor contribution in the considered regime. In this case the plane wave laser field could not modify the electron dynamics after the laser-electron interaction \cite{Salamin2002} (see Fig.~\ref{fig3}(c)) because the electron cannot absorb laser photons from a plane wave field due to the momentum conservation. However, in the focused laser field the electron can absorb laser photons due to an additional ponderomotive momentum transfer to the field which stems from the transverse gradient of the laser field and, therefore, the laser focusing effect induces the electron reflection. Since $\gamma\ll\xi/2$ for the applied parameters, the electrons are reflected at the rising edge of the laser pulse (at $\eta/2\pi\approx 7$, see Fig.~\ref{fig3}(b)), and with different CEPs the electron dynamics is apparently different. Furthermore, we follow the tracks of a group of sample electrons near the laser polarization plane ($y=0$). The initial coordinate distribution of the sample electrons are given in Fig.~\ref{fig3}(d). The electrons initially distributed along the transverse profile of the laser field (e.g., marked in blue, red, yellow, and orange) are subjected to different transversal gradient of the laser field $d I/d x$, which determines the final angular spread, see Figs.~\ref{fig3}(e) and \ref{fig3}(f). For the electrons initial near the intensity peak (marked in green and purple), the intensity transverse gradient is rather small and, therefore, the final angular spread is small. Consequently, electron density peaks are formed  mostly by those electrons initially near the beam center. In the focused laser fields with different CEPs, the electrons experience different laser fields, and their dynamics and final momentum distribution are rather different, see Figs.~\ref{fig3}(e) and \ref{fig3}(f) (also Figs.~\ref{fig2}(a) and \ref{fig2}(b)). Therefore, the electron-density peak would be sensitively governed by the CEP.

 \begin{figure}[t]
	 \includegraphics[width=0.95\linewidth]{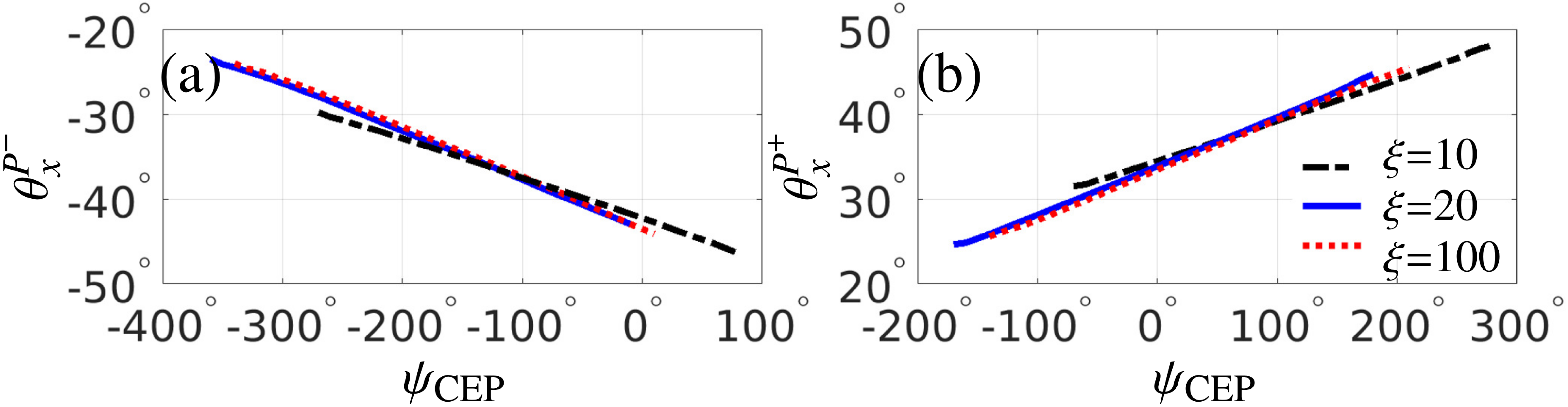}
	\caption{\label{fig4} (a), (b) The variations of $\theta_x^{P^{-}}$ and $\theta_x^{P^{+}}$ with respect to the CEP, respectively. The black-dash-dotted, blue-solid, and red-dotted curves represent the cases of $\xi=10$, 20, and 100, respectively, and the corresponding electron energies $\varepsilon_i=$ 45 keV, 100 keV, and 2 MeV, respectively. Other parameters are the same as in Fig.~\ref{fig2}.}
\end{figure}

We further investigate the impact of the laser and electron parameters on the CEP determination. 
The dependence on the laser intensity is illustrated in Fig.~\ref{fig4}. For $\xi=10$, 20, and 100, $\Delta \theta_x^{P^{-}}=$ Max($\theta_x^{P^{-}}$)-Min($\theta_x^{P^{-}}$) = 16.65$^{\circ}$, 20.05$^{\circ}$, and 20.08$^{\circ}$, respectively, and $\Delta \theta_x^{P^{+}}=$ 16.62$^{\circ}$, 20.05$^{\circ}$, and 19.93$^{\circ}$, respectively. The gradient increases with $\xi$ as $\xi\lesssim 20$ and becomes stable as $\xi\gtrsim20$. 
The CEP resolutions are approximately $0.12^{\circ}$, $0.1^{\circ}$, and $0.1^{\circ}$, respectively. Thus, the CEP resolution is inversely  propotional to the $\xi$ parameter as $\xi\lesssim20$. For instance, it  is about $0.2^\circ$ as $\xi=5$ \cite{supplemental}. As $\xi$ decreases, the required $\varepsilon_i$ decreases as well to satisfy $\gamma\ll \xi/2$.   However, $\xi$ is limited from below by the requirement of the relativistic interaction, which is a prerequisite for the applied regime. 

For realistic experimental conditions of petawatt laser pulses, the laser-energy detection can achieve an uncertainty of about 1.5\% \citep{Sung2017} (usually better than 1\% for the-state-of-the-art terawatt laser pulses  \cite{Adolph2017}). We find that as a $4\%$ (1\%) laser-energy (or laser-intensity) uncertainty is taken into account, the CEP detection has merely an about 2.8\% (1.1\%) uncertainty \cite{supplemental}, which is better than recent experimental achievement for few-cycle multi-terawatt laser systems \cite{Adolph2017}. Moreover, the angle-resolved electron number can be employed to determine the CEP as well, see analysis in \cite{supplemental}.

\begin{figure}[t]
	\includegraphics[width=0.95\linewidth]{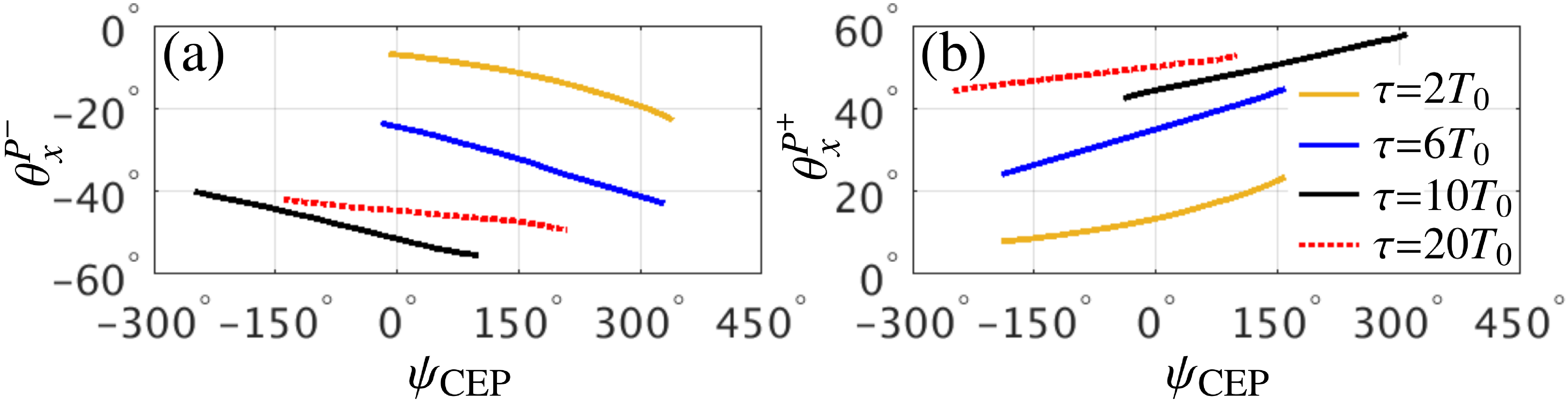}
	\caption{\label{fig5} 	(a), (b) The variations of $\theta_x^{P^{-}}$ and $\theta_x^{P^{+}}$ with respect to the CEP, respectively. The yellow, blue, and black curves represent the cases of the focused Gaussian laser pulse with $\tau=2$, 6, and 10$T_0$, respectively, and the red curves  show the case of a flat-top Gaussian laser pulse with $\tau=20 T_0$. Other parameters are the same as in Fig.~\ref{fig2}}
  \end{figure}

  Commonly, the CEP effects in the laser-matter interaction are  negligible in the  case of long laser pulses ($\tau\gtrsim 6T_0$). However, the specific CEP effect in the relativistic regime of interaction considered in this Letter is still significant for multicycle laser pulses, because it originates from the field asymmetry on the rising edge of the laser pulse. The dependence of the proposed CEP signature on the laser pulse duration is analyzed in Fig.~\ref{fig5}. We compare the variations of $\theta_x^{P^{-}}$ and $\theta_x^{P^{+}}$ with respect to  the CEP for Gaussian laser pulses with $\tau=2 T_0$ (yellow), 6$T_0$ (blue), and 10$T_0$ (black), respectively. As the laser pulse duration increases, the gradients of the variations of $\theta_x^{P^{-}}$ and $\theta_x^{P^{+}}$ decrease slightly. The CEP resolutions for all cases are close to $0.1^{\circ}$.  Moreover, the case of a flat-top Gaussian laser pulse with a pulse envelop of exp($-t^4/\tau^4$) and $\tau=20 T_0$ (red) is shown as well. Since this method detects the CEP via the electron reflection at the rising edge of the laser pulse, it works well also for the long flat-top laser pulses. The CEP resolution for the given parameters is $\sim 0.3^{\circ}$. 
  
  The role of the initial kinetic energy $\varepsilon_i$ of the electron bunch is investigated 
as well, see~\cite{supplemental}. As $\varepsilon_i$ increases from 1 keV to 1 MeV with a constant $\xi=20$, the CEP resolution reduces from $\sim 0.1^{\circ}$ to $\sim0.25^{\circ}$ slowly, and the optimal $\varepsilon_i$ fulfills the condition $\gamma\sim\xi/100$.  Note that in the considered scheme of  CEP determination, the laser intensity and the electron energy are both limited from above such that the stochastic effects of radiation are negligible, i.e., $\chi\sim 10^{-6}\xi\gamma \ll 1$. Otherwise  the CEP signatures vanish completely, see \cite{supplemental}.

In conclusion, we have investigated a new efficient method for the CEP determination of long relativistic tightly-focused laser pulses via the electron dynamics. As the conditions of $\gamma\ll\xi/2$ and $\chi\ll1$ are satisfied for the applied parameters, the electrons can be reflected by the rising edge of the driving laser pulse because of the laser focusing effects. In the laser polarization direction, the electron momenta distribute asymmetricly, and two peaks emerge in the angular spectra of the momenta, which could senstively reveal the CEP. This method is shown to be robust with the laser and electron parameters and is applicable for relativistic laser pulses of $I_0\gtrsim10^{19}$~W/cm$^2$ and the pulse duration up to 10 cycles (up to 20 cycles for the flat-top laser pulses), which can be successfully applied in the current standard strong-field laser facilities. The CEP resolution can achieve about  $0.1^{\circ}$, one order of magnitude higher than that of the CEP determination via the x-ray radiation \cite{Li2018}, since the diffraction limitation of an electron is much smaller than that of a photon.

In the near future, with fast  development of ultraintense lasers and the broad research interests on strong laser physics, the accurate determination of CEP of relativistic laser pulses are expected to be in great demand. Interesting point is that the considered relativistic CEP effect arises in the rising edge of the laser pulse which allows for  CEP measurement of relativistic laser pulses.

This work is supported by the  Science Challenge Project of China (Nos. TZ2016005, TZ2016099), the National Key Research and Development Program of China (Grant No. 2018YFA0404801), the National Natural Science Foundation of China (Grants Nos. 11874295, 11804269, U1532263, 11520101003, and 11861121001), and the Laser Fusion Research Center Funds for Young Talents (Grant No. RCFPD2-2018-4).

\bibliography{Manuscript}

\end{document}